\documentclass[%
 reprint,
 superscriptaddress,
 showpacs,
 amsmath,amssymb,
 aps,prl,
floatfix
]{revtex4-1}

\usepackage{graphicx}
\usepackage{dcolumn}
\usepackage{bm}

\begin{document}

\preprint{preprint number}

\title{A Search for Invisible Axion Dark Matter with the Axion Dark Matter Experiment}

\author{N. Du}%
  \affiliation{University of Washington, Seattle, WA 98195, USA}
  \author{N. Force}
  \affiliation{University of Washington, Seattle, WA 98195, USA}
\author{R. Khatiwada}
  \affiliation{University of Washington, Seattle, WA 98195, USA}
\author{E. Lentz}
  \affiliation{University of Washington, Seattle, WA 98195, USA}
\author{R. Ottens}
  \affiliation{University of Washington, Seattle, WA 98195, USA}
 \author{L. J Rosenberg}%
  \affiliation{University of Washington, Seattle, WA 98195, USA}
  \author{G. Rybka}%
  \email[Correspondence to:]{grybka@uw.edu}
  \affiliation{University of Washington, Seattle, WA 98195, USA}

\author{G. Carosi}
\affiliation{Lawrence Livermore National Laboratory, Livermore, CA 94550, USA}
\author{N. Woollett}
\affiliation{Lawrence Livermore National Laboratory, Livermore, CA 94550, USA}

\author{D. Bowring}
  \affiliation{Fermi National Accelerator Laboratory, Batavia IL 60510, USA}
\author{A. S. Chou} 
  \affiliation{Fermi National Accelerator Laboratory, Batavia IL 60510, USA}
\author{A. Sonnenschein} 
  \affiliation{Fermi National Accelerator Laboratory, Batavia IL 60510, USA}
  \author{W. Wester} 
  \affiliation{Fermi National Accelerator Laboratory, Batavia IL 60510, USA}
  
\author{C. Boutan}
  \affiliation{Pacific Northwest National Laboratory, Richland, WA 99354, USA}
  \author{N. S. Oblath}
  \affiliation{Pacific Northwest National Laboratory, Richland, WA 99354, USA}

\author{R. Bradley}
\affiliation{National Radio Astronomy Observatory, Charlottesville, VA 22903, USA}

\author{E. J. Daw}
\affiliation{University of Sheffield, Sheffield UK}

\author{A. V. Dixit}
\affiliation{University of Chicago, IL 60637}

\author{J. Clarke}
  \affiliation{University of California, Berkeley, CA 94720, USA}
\author{S. R. O'Kelley}
  \affiliation{University of California, Berkeley, CA 94720, USA}
 
 \author{N. Crisosto}
  \affiliation{University of Florida, Gainesville, FL 32611, USA}
\author{J.~R.~Gleason}
  \affiliation{University of Florida, Gainesville, FL 32611, USA}
\author{S. Jois}
  \affiliation{University of Florida, Gainesville, FL 32611, USA}
 \author{P. Sikivie}
  \affiliation{University of Florida, Gainesville, FL 32611, USA}
\author{I. Stern}
  \affiliation{University of Florida, Gainesville, FL 32611, USA}
\author{N.S. Sullivan}
  \affiliation{University of Florida, Gainesville, FL 32611, USA}
\author{D.B Tanner}
  \affiliation{University of Florida, Gainesville, FL 32611, USA}
  \collaboration{ADMX Collaboration}\noaffiliation

\author{G. C. Hilton}
\affiliation{National Institute of Standards and Technology, Boulder, Colorado 80305 USA}

\date{\today}

\begin{abstract}
This Letter reports results from a haloscope search for dark matter axions with masses between 2.66 and 2.81 $\mu$eV. The search excludes the range of axion-photon couplings predicted by plausible models of the invisible axion.  This unprecedented sensitivity is achieved by operating a large-volume haloscope at sub-kelvin temperatures, thereby reducing thermal noise as well as the excess noise from the ultra-low-noise SQUID amplifier used for the signal power readout.  Ongoing searches will provide nearly definitive tests of the invisible axion model over a wide range of axion masses.

\end{abstract}


\maketitle


  
Axions are particles predicted to exist as a consequence of the Peccei-Quinn solution to the strong-CP problem \cite{Peccei:1977hh,Weinberg:1977ma,Wilczek:1977pj}, and could account for all of the dark matter in our Universe \cite{Abbott:1982af,Preskill:1982cy,PhysRevLett.50.925}.  
While there exist a number of mechanisms to produce axions in the early universe \cite{Abbott:1982af,DINE1983137,Sikivie:2006ni,PhysRevD.78.083507} that allow for a wide range of dark matter axion masses, current numerical and analytical studies of QCD typically suggest a preferred mass range of 1--100 $\mu$eV for axions produced after cosmic inflation in numbers that saturate the Lambda-CDM cold dark matter density \cite{Bonati2016,PhysRevD.92.034507,Borsanyi2016,PhysRevLett.118.071802,PhysRevD.96.095001}.  The predicted coupling between axions and photons is model-dependent; in general, axions with dominant hadronic couplings as in the KSVZ model \cite{Kim:1979if,Shifman:1979if} are predicted to have an axion-photon coupling roughly 2.7 times larger than that of the DFSZ model\cite{Dine:1981rt,Zhitnitsky:1980tq}.  Because the axion-photon coupling is expected to be very small, ${\cal O}(10^{-17}-10^{-12}$~GeV$^{-1})$ over the expected axion mass range, these predicted particles are dubbed {\it invisible} axions \cite{Abbott:1982af}. 

The most promising technique to search for dark matter axions in the favored mass range is the axion haloscope \cite{Sikivie:1983ip}, consisting of a cold microwave resonator immersed in a strong static magnetic field.  In the presence of this magnetic field, the ambient dark matter axion field produces a volume-filling current density, oscillating at frequency $f = {\cal E}/h$, where $\cal E$ is the total energy consisting mostly of the axion rest mass with a small kinetic energy addition. When the resonator is tuned to match this frequency, the current source delivers power to the resonator in the form of microwave photons which can be detected with a low-noise microwave receiver.
To date, a number of axion haloscopes have been implemented. All had noise levels too high to detect the QCD axion signal  \cite{DePanfilis:1987dk,Wuensch:1989sa,Hagmann:1990tj,PhysRevLett.80.2043,PhysRevD.64.092003,1538-4357-571-1-L27,PhysRevD.69.011101,Asztalos:2009yp,PhysRevD.94.082001,Sloan201695,PhysRevLett.118.061302} in an experimentally realizable time.  Previous versions of the Axion Dark Matter eXperiment (ADMX)   \cite{PhysRevD.64.092003,1538-4357-571-1-L27,PhysRevD.69.011101,Asztalos:2009yp,PhysRevD.94.082001,Sloan201695} achieved sensitivity to the stronger KSVZ couplings in the 1.91-3.69 $\mu$eV mass range.  ADMX has since been improved to utilize a dilution refrigerator to obtain a significantly lower system noise temperature, drastically increasing its sensitivity.  We present here results from the first axion experiment to have sensitivity to the more weakly coupled DFSZ axion dark matter in the micro-eV mass range.


The Generation 2 ADMX experiment consists of an 136-liter cylindrical, copper-plated microwave cavity, placed inside a superconducting magnet.  
The geometry and thus frequency of the cavity is changed by means of two copper-plated tuning rods, extending the length of the cavity interior, which can be moved from near the center to the perimeter in very small increments.
The cavity, magnet, and tuning system are described in more detail in Refs.~\cite{Asztalos:2009yp,ASZTALOS201139}.  If the TM$_{010}$ cavity resonant mode radio-frequency (RF) overlaps with the frequency of photons from dark matter axion conversion, power is expected to develop in the cavity in excess of thermal noise:
\begin{widetext}
\begin{equation}
P_{\mathrm{axion}}=1.9\times 10^{-22} \mathrm{W} \left(\frac{V}{136 \mathit{\ l}}\right) \left(\frac{B}{6.8 \mathrm{\ T}}\right)^2 \left(\frac{C}{0.4}\right)\left(\frac{g_\gamma}{0.97}\right)^2\left(\frac{\rho_\mathrm{a}}{0.45\mathrm{\ GeV\,cm^{-3}}}\right)\left(\frac{f}{650 \mathrm{\ MHz}}\right)\left(\frac{Q}{50,000}\right).
\label{power_equation}
\end{equation}
\end{widetext}
Here $V$ is the cavity volume, $B$ is the magnetic field, $C$ is a form factor representing the overlap between the microwave electric field and the static magnetic field, $g_\gamma$ is the model dependent part of the axion-photon coupling with a value of -0.97 and 0.36 for the KSVZ and DFSZ benchmark models, respectively, $\rho_\mathrm{a}$ is the axion dark matter density at the Earth's location, $f$ is the frequency of the photons from axion conversion, and $Q$ is the loaded cavity quality factor.  This power has been scaled to typical experimental parameters for the results reported here. 

Power in the the TM{$_{010}$ mode of the cavity is extracted with a critically coupled antenna, passed through the RF chain shown in Fig.~\ref{fig:rfchain}, and amplified by a voltage-tunable Microstrip SQUID Amplifier (MSA) \cite{doi:10.1063/1.121490,doi:10.1063/1.125383} located in a magnetic field-free region.  The signal is then passed through a cryogenic Heterostructure Field-Effect Transistor (HFET) amplifier, and mixed with a local oscillator to center the cavity resonant frequency at the 10.7 MHz intermediate frequency for further processing and analysis. The operation of an MSA in an axion experiment is described in more detail in Ref.~\cite{ASZTALOS201139}.  The signal is digitized and the voltage-time series is converted into a power-frequency spectrum over a 25-kHz bandwidth, which roughly matches the bandwidth of the cavity. 

The signal-to-noise ratio ($\mathrm{SNR}$) of an axion signal power to thermal noise power is of paramount importance in exploring axion masses rapidly.  It is given by \cite{doi:10.1063/1.1770483}
\begin{equation}
\mathrm{SNR} = \left(P_{\mathrm{axion}}/{kT_{\mathrm{system}}}\right)\left(t/b\right)^{\frac{1}{2}},
\end{equation}
where $k$ is Boltzmann's constant, $T_{\mathrm{system}}$ is the sum of the physical temperature of the cavity and the noise temperature of the receiver, $t$ is the time spent integrating at a particular frequency, and $b$ is the bandwidth of the axion signal, set by the local axion velocity distribution.

The cavity and MSA are cooled by a  dilution refrigerator to minimize thermal background and excess thermal noise from the amplifier. The refrigerator has a cooling power of 800 $\mu$W at 100 mK.  The cavity temperature, as measured by ruthenium oxide thermometers
throughout the run, was typically 150 mK, with the MSA temperature measured to be about 300 mK due to additional heat in the vicinity of the MSA.  The expected contribution to the system noise of the MSA is bounded from below by the standard quantum limit (30 mK at the frequencies reported in this paper) and is typically near half of its physical temperature.

The power in the receiver was calibrated from the temperature sensors by comparing power measured on and off the cavity resonance.  Off resonance, the MSA amplified primarily Johnson noise from attenuator A4 in Fig.~\ref{fig:rfchain} with a temperature around 300 mK.  On-resonance the MSA amplified primarily the blackbody radiation from the cavity with a temperature around 150 mK.  The noise power spectrum was fit with a model of the RF chain  (Fig.~\ref{fig:noise}) to determine the total noise temperature.  More detail is provided in the supplementary material and reference \cite{friis1944}.

\begin{figure}
\begin{centering}
\includegraphics[width=0.5\textwidth]{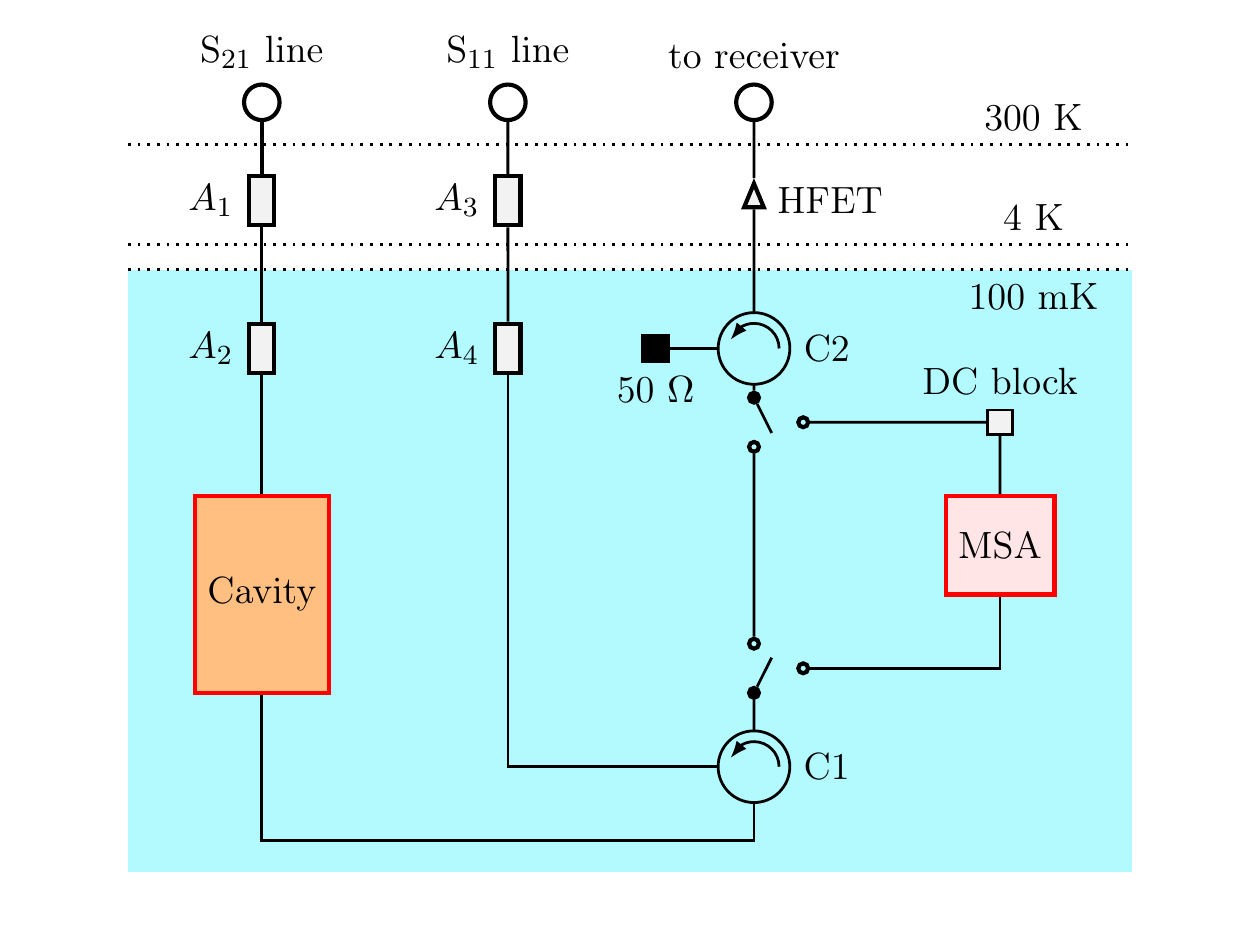}
\caption{ADMX Cryogenic RF chain. C1 and C2 are circulators, MSA is the Microstrip SQUID Amplifier, A1-A4 are attenuators, and HFETs, are the cooled transistor amplifiers. In the data-taking configuration, the output of the cavity is sent via C1 to be amplified by the MSA, via C2 to be amplified further by the HFETs, and to the receiver for mixing to 10.7 MHz, further amplification and digitization of the power from the cavity.   Network analyzer transmission (S$_{21}$) and reflection measurements (S$_{11}$) are made before each digitization. \label{fig:rfchain}}
\end{centering}
\end{figure}


The results  reported  here are based on data acquired between January 18, 2017 and June 11, 2017.  The data acquisition and analysis procedure are similar to those described in Ref.~\cite{Asztalos:2009yp} and are summarized here.  A single cycle of data acquisition consists of a small frequency step via the physical positioning of the tuning rods, measuring the TM$_{010}$ mode frequency and loaded $Q_L$ via $S_{21}$ transmitted power using a network analyzer, measuring the coupling to the mode with $S_{11}$ reflected power, also with a network analyzer, and then digitizing the power coming out of the cavity for a period of 100 s in a bandwidth of 25 kHz centered on the TM$_{010}$ resonant frequency.  

Our results reported are based on 78,958 spectra each 25 kHz wide.  Under optimal experimental conditions, a typical frequency bin achieved desired sensitivity when measured by 20 overlapping spectra (see Fig.~\ref{fig:datanalysis}).  Periodically during operation, we evaluated the expected sensitivity to an axion signal and more scans were added to compensate for low-sensitivity regions due to a varying noise temperature or tuning speed.
With sufficient raw data collected, a preliminary analysis was performed to identify spectral features consistent with an excess power from an axion signal.

Analysis consisted of first generating a power spectrum from each 100-s digitization with a 96-Hz bin size, following the procedure outlined in Ref.~\cite{Brubaker:2017rna}.  The receiver transfer function spectral shapes were removed with a Savitsky-Golay filter (length 121 and polynomial order 4) to 95\% of the least-deviant power bins, thus removing structures much broader than axion signals.  If the standard deviation of the 95\% least deviant points was more than 20\% above the expected standard deviation for white noise, the background fit was declared poor and the data were removed from further analysis steps.  
The power was scaled to the known system noise and weighted by $Q_L$ to produce a measurement of power excess in each bin attributable to an axion signal.  
This power excess was then optimally filtered by convolving with two astrophysically-motivated axion signal shapes: first, the boosted Maxwell-Boltzmann predicted by the standard halo model (SHM) of axion dark matter \cite{PhysRevD.42.3572} (which has a linewidth of roughly 700 Hz at the frequencies reported here), and second by the N-body derived lineshape described in Ref.~\cite{0004-637X-845-2-121} (with a linewidth nearly half that of the Maxwell-Boltzmann), each model yielding the excess power attributable to an axion of a given mass.  
When the data were statistically consistent with no axion signals being present, the signal power measurement and uncertainty could be used to set an upper limit on axion-photon coupling using Eq.~\ref{power_equation}.  Frequencies at which the power was in excess of $3\sigma$ above the mean were labeled as ``candidates'' and flagged for rescan and further analysis. 

Candidate signals were rescanned to equivalent sensitivity to measure their persistence. If the excess power persisted in any of the candidates, then a second and longer rescan was performed at the candidate signal frequencies for $3$ times as long to improve local sensitivity and candidate significance.  Any frequencies where excess power persisted following the second rescan were analyzed individually for the possibility of RF interference.

We tested the performance of the analysis by imposing software-simulated axion signals on to the raw power spectra.  We injected 25,000 software-simulated signals into the dataset with couplings varying between DFSZ and 10 times KSVZ, ran through the analysis process, and evaluated the resulting candidate power to determine the systematic uncertainty associated with the background subtraction.  Figure~\ref{fig:datanalysis} shows the effect of injected signals in both the background-subtracted spectra and the final filtered and combined spectrum.

In addition to the uncertainty introduced by the analysis procedure, there are systematic uncertainties in the product of axion-photon coupling constant and dark matter density from the temperature measurement, noise calibration, Q measurement, and numerical modeling of the form factor in Eq.~\ref{power_equation}, shown in Table~\ref{tab:systematics}.  However, the sensitivity of the results reported here is restricted primarily by the statistics of the finite observation time at each frequency.

\begin{figure}
\begin{centering}
\includegraphics[width=0.5\textwidth]{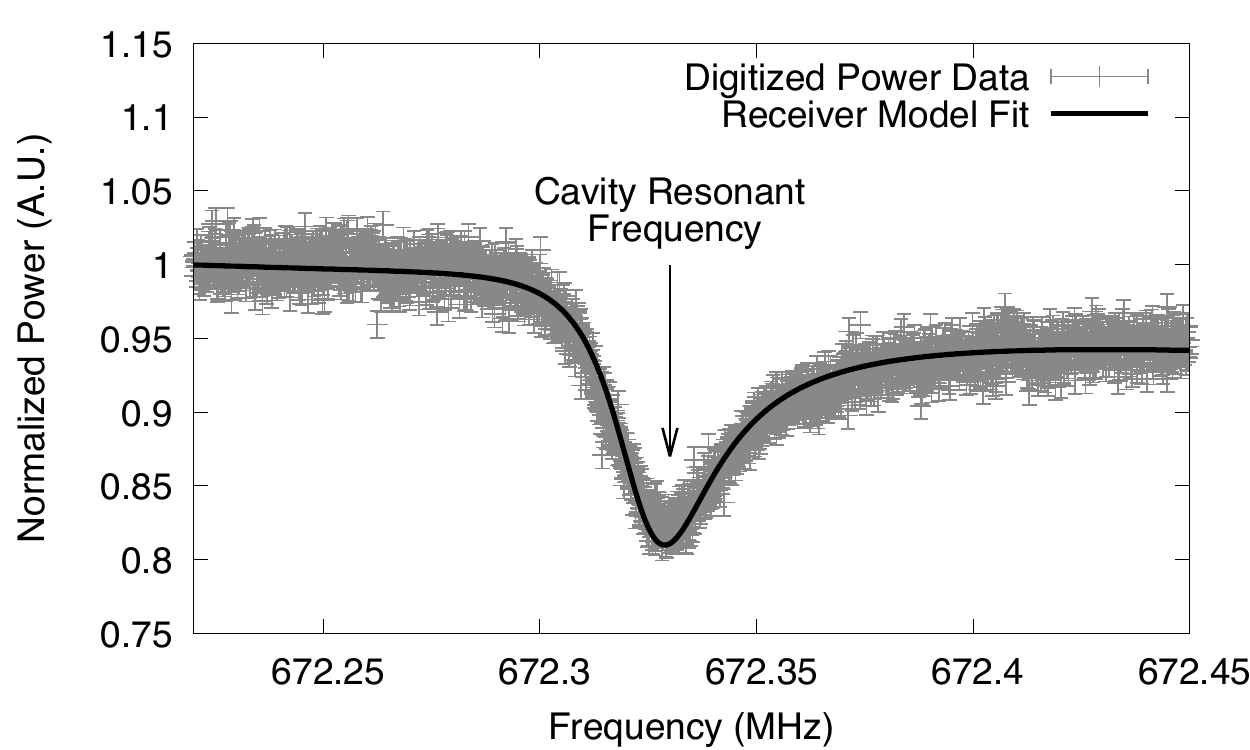}
\caption{One of the power measurements used to calibrate the system noise temperature.  Off resonance, the power is the sum of the 300~mK physical temperature of an attenuator and the receiver noise temperature.  On resonance, the power is the sum of the 150~mK physical temperature of the cavity and the receiver noise temperature.  The noise power on versus off resonance acts as an effective hot-cold load, with the physical temperatures measured with sensitive thermometers.  The asymmetry of the shape is a result of interactions between RF components, as described in the supplementary material\label{fig:noise}}
\end{centering}
\end{figure}

\begin{figure}
\begin{centering}
\includegraphics[width=0.5\textwidth]{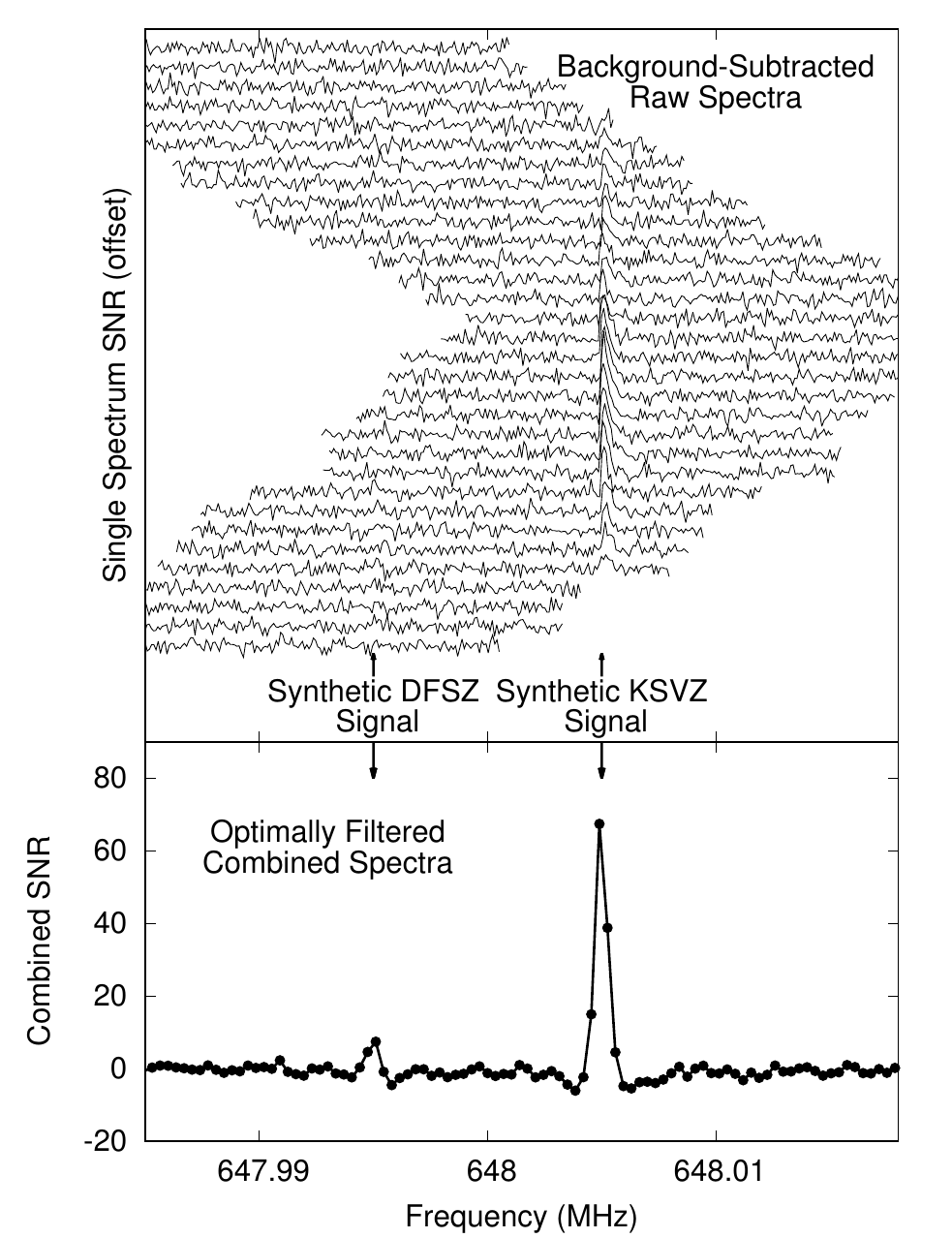}
\caption{\textit{Upper figure}: Series of background-subtracted single scans with synthetic axion signals with the N-body inspired signal shape \cite{0004-637X-845-2-121}, one at KSVZ coupling and one at DFSZ coupling.  The KSVZ signal is easily visible in these individual spectra; the DFSZ signal, being a factor of 7 smaller, is not.  \textit{Lower figure}: Same data after the individual scans have been optimally filtered and combined.  Both KSVZ and DFSZ signals are visible with high SNR. \label{fig:datanalysis}}
\end{centering}
\end{figure}

\begin{table}

\begin{tabular}{|l|c|}
\hline
\textbf{Source} & \textbf{$g_\gamma^2 \rho_\mathrm{a}$ Uncertainty} \\
\hline
Temperature Sensor Calibration & 7.1\%\\
System Noise Calibration & 7.5\%\\
Quality Factor Measurement & 2.2\%\\
Background Subtraction & 4.6\%\\
Form Factor Modeling & 6.0\%\\
\hline
\textbf{Total} & 13\%\\
\hline
\end{tabular}
\caption{Primary sources of systematic uncertainty.  The form factor uncertainty varies somewhat with frequency; the value at 655 MHz is shown here.  The combined effect of systematic uncertainty on the exclusion bounds is shown as the width of the lines in Fig.~\ref{fig:exclusion}. \label{tab:systematics}}
\end{table}


In the range 645--680 MHz, no statistically significant signals consistent with axions were found.  There were two candidates that persisted after the rescan procedure, but a measurement of the external background radio interference at the experimental site found the identical external radio signals at the candidate frequencies.  They are thus excluded from our limits.  We are therefore able to produce a 90\% upper confidence limit on the axion photon coupling using all of the data acquired, for the Maxwellian and N-body astrophysical models, shown in Fig.~\ref{fig:exclusion}.   We are able to exclude both DFSZ axions distributed in the isothermal halo model that make up 100\% of dark matter with a density of 0.45 GeV/cm$^3$ and DFSZ axions with the N-body inspired lineshape and the predicted density of 0.63 GeV/cc between the frequencies 645 and 676 MHz.  
This result is a factor of 7 improvement in power sensitivity over previous results and the first time an axion haloscope has been able to exclude axions with DFSZ couplings.

\begin{figure*}
\includegraphics[width=0.95\textwidth]{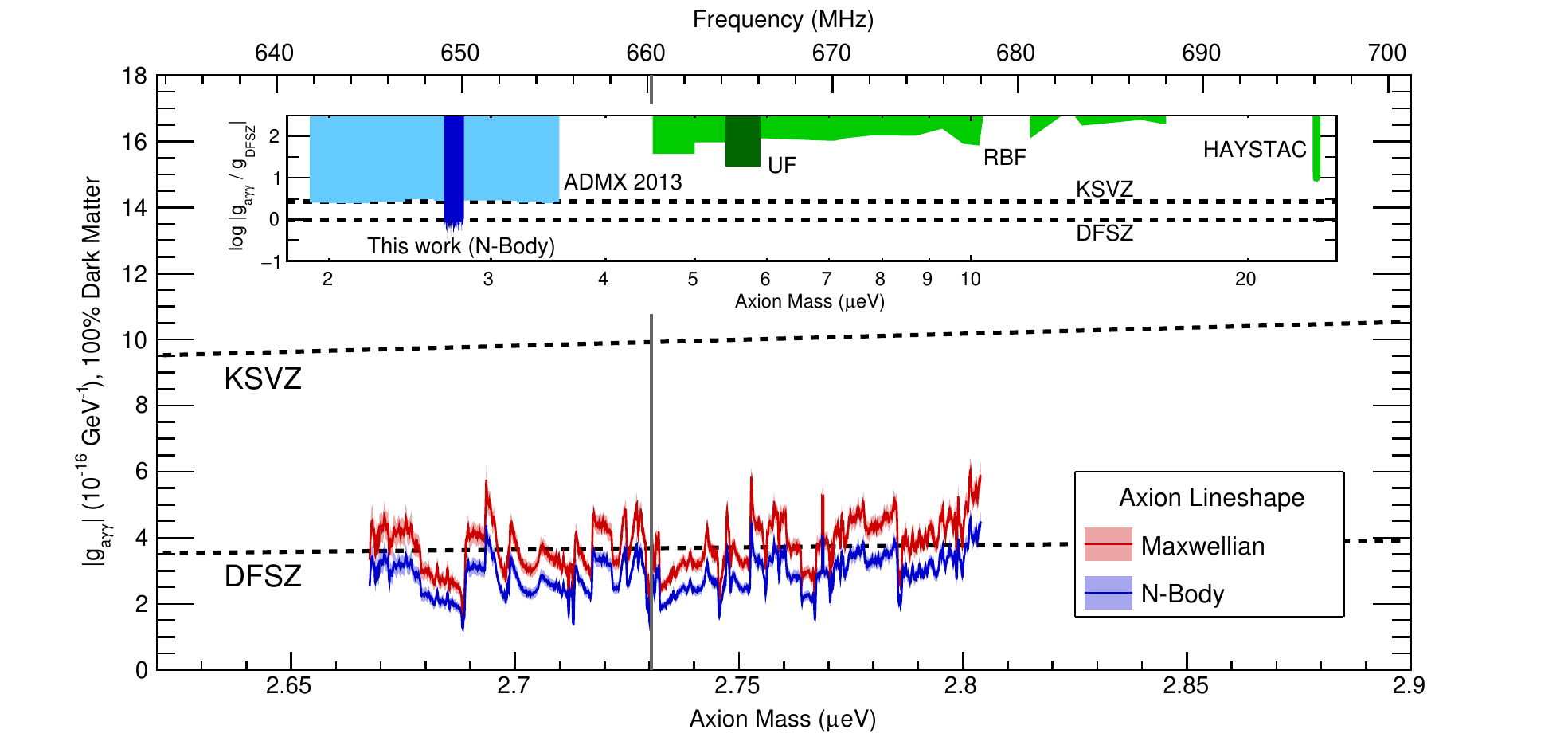}
\caption{90\% upper confidence excluded region of axion mass and photon coupling $g_{a\gamma\gamma}$. The red line indicates the limit on axion-photon coupling with the boosted Maxwel-Boltzman lineshape from the isothermal halo model~\cite{PhysRevD.42.3572}, while the blue like indicates the limit with the N-body inspired signal~\cite{0004-637X-845-2-121}.  Colored regions indicate systematic uncertainty range.  The region 660.16 to 660.27 MHz, marked by the grey bar, was vetoed due to RF interference as described in the text.  The inset shows the results in the context of other haloscope searches.   \label{fig:exclusion}}
\end{figure*}


ADMX has achieved a factor of 7 improvement in its already world-leading sensitivity to ultralow signal power levels. It is the only operating experiment able to probe the DFSZ GUT coupling for the invisible axion that has long been the goal of the axion search community.  Data from this period of initial operations have now excluded these models over a range of axion masses.  A much larger range of masses will be probed in future runs; we expect to operate the apparatus at lower temperature and with a greater magnetic field, enabling higher scan speeds.  A recent engineering run of the apparatus (with some electronics removed) achieved cavity temperatures lower than reported in this paper, while the magnet in earlier ADMX runs\cite{PhysRevD.64.092003} was operated at 7.6~T compared to the typical field of 6.8~T for the results reported here. Together, these improvements could increase the SNR by a factor of two or shorten the measurement time by a factor of four.  Coverage of masses up to 40~$\mu$eV (10~GHz) is envisioned by further augmenting the signal power by combining the outputs of multiple co-tuned cavity resonators inside the current magnet.  A discovery could occur at any point during this process, and a confirmation with independent data can be quickly achieved given the short integration times needed to re-acquire the signal at the correct cavity tuning. The signal, once found, will \textit{always} be there.  This experiment heralds a new era of ultra-sensitive probes of low mass axionic dark matter, the discovery of which would also confirm the Peccei-Quinn solution\cite{Peccei:1977hh,Weinberg:1977ma,Wilczek:1977pj} to the long-standing CP problem of the strong interaction.

This work was supported by the U.S. Department of
Energy through Grants No. DE-SC0009723,No. DE-SC0010296, No. DE-FG02-96ER40956, No. DEAC52-07NA27344, 
and No. DE-C03-76SF00098. Fermilab is a U.S. Department of Energy, Office of Science, HEP User Facility. Fermilab is managed by Fermi Research Alliance, LLC (FRA), acting under Contract No. DE-AC02-07CH11359.
Additional support was provided by the Heising-Simons
Foundation and by the Lawrence Livermore National
Laboratory and Pacific Northwest National Laboratory LDRD programs.

\bibliographystyle{apsrev4-1}
\bibliography{2017admx_main}


\end{document}